\documentclass{article}
\usepackage{epsfig}
\usepackage{amsmath}
\usepackage{amssymb}
\usepackage[hang,nooneline,scriptsize]{subfigure}

\usepackage[dvips]{color}
\usepackage[normalem]{ulem}

\newcommand{\ee}{\end{equation}}
\newcommand{\eea}{\end{eqnarray}}
\newcommand{\be}{\begin{equation}}
\newcommand{\bea}{\begin{eqnarray}}

\begin{document}

\title{\LARGE \bf Proca Q-Tubes and their Coupling to Gravity}
  \author{
  \large  Y. Brihaye$^a$ $\:${\em and}$\:$
  Y. Verbin$^b$ \thanks{Electronic addresses: yves.brihaye@umons.ac.be; verbin@openu.ac.il } }
 \date{ }
   \maketitle
    \centerline{$^a$ \em Physique Th\'eorique et Math\'ematiques, Universit\'e de Mons,}
   \centerline{\em Place du Parc, B-7000  Mons, Belgique}
     \vskip 0.4cm
   \centerline{$^b$ \em Department of Natural Sciences, The Open University
   of Israel,}
   \centerline{\em Raanana 43107, Israel}

 \maketitle
\begin{abstract}
The  Einstein-Proca system is studied in the case of a complex vector field
self-interacting through an appropriate potential with a global U(1) symmetry. The corresponding equations for a static, cylindrically symmetric metric and matter fields are then constructed and solved.
In the probe limit (no gravity), it is shown that the equations admit at least two classes of regular solutions
distinguished by the asymptotic behavior of the matter fields. One of these classes corresponds
to lumps of vector fields localized in a cylindrical region,  we naturally call these solutions ``Proca  Q-tubes''.
They constitute the cylindrical counterparts of spherical Proca Q-balls constructed recently; they
can be  characterized by  finite mass and charge per unit length of the tube.
The domain of existence of these Proca Q-tubes  with respect to the coupling constants
determining the potential is studied in detail. Finally, the gravitating Proca Q-tubes are constructed and studied.
\end{abstract}

\maketitle

\medskip
 \ \ \ PACS Numbers: 04.70.-s,  04.50.Gh, 11.25.Tq

\section{Introduction}\label{Introduction}
\setcounter{equation}{0}

In the recent couple of years a substantial number of papers appeared, studying and analyzing non-topological solitons ``made of'' self-interacting vector fields \cite{Loginov2015,BritoEtAl2015,Landea-Garcia2016,Duarte-Brito2016} whose dynamics is set by the following Lagrangian density (we use the ``mostly minus'' metric sign convention):
\be \label{matterLag}
{\cal L} = -\frac{1}{4} F^*_{\mu\nu}F^{\mu\nu} - U(A^*_\mu A^\mu)
\ee
where as usual $A_{\mu}$ is the vector potential and $F_{\mu\nu}=\partial_{\mu}A_{\nu}-\partial_{\nu}A_{\mu}$ is the corresponding field strength - although they may be complex. The potential  may contain only a mass term, $U(A^*_\mu A^\mu)=-m^2 A^*_\mu A^\mu /2$ as in the original Proca Lagrangian \cite{Proca1936,Proca1937,Poenaru2005}, in which case localized solutions exist only with the help of gravity. These structures are known as Proca stars \cite{BritoEtAl2015,Landea-Garcia2016,Duarte-Brito2016}. There exist also some earlier studies \cite{Obukhov1999,Vuille2002} from around the year 2000.

On the other hand, a higher order polynomial like
\be \label{potential}
U(\psi) = -\frac{m^2}{2} \psi -  \frac{\lambda}{4} \psi^2 - \frac{\nu}{6} \psi^3  \,\,\,\,\,\, , \,\,\,\psi=A^*_\mu A^\mu
\ee
may support localized solutions even in flat space without invoking gravity \cite{Loginov2015}. These may be named Proca Q-balls in flat space or Proca Q-stars for the self-gravitating version, reflecting the strong similarity with the analogous solutions of scalar fields \cite{FriedbergLeeSirlin1976,Coleman1985,Lynn1988,Jetzer1989,Lee-Pang1991}. The reason for the sign difference with respect to the scalar field potential is the ``mostly minus'' metric sign convention that we use. However, with this choice, the vector localized solutions require $\lambda>0$ as in the scalar case (while $\nu>0$ in both cases).

All kinds of self-interaction are covered by the following action ($R$ is Ricci scalar and $G$ Newton's constant):

\be \label{totalAction}
S = \int d^4 x \sqrt{- g} \left (\frac{R}{16 \pi G} -\frac{1}{4} F^*_{\mu\nu}F^{\mu\nu} - U(A^*_\mu A^\mu) \right )
\ee

In spite of the strong similarity, we stress already here an important difference between the vector and scalar
potential functions. Unlike the scalar case, the potential $U(A^*_\mu A^\mu)$ is unbounded from below because of the non-definite spacetime norm. However, the energy (or mass) density turns out still to be bounded from below for a certain range of parameters as we will see in the next section.
At any rate, the potential function breaks explicitly the local $U(1)\times U(1)$ symmetry of the pure (double) Maxwell theory (the kinetic term), and leaves a global $U(1)$ only. A non-Abelian generalization has been studied as well \cite{Ponglertsakul-Winstanley2016}.

Another aspect of these theories that was studied quite extensively is black hole solutions that were found to exist exhibiting a vector hair in either the Abelian \cite{HerdeiroEtAl2016,Fan2016} or non-Abelian case \cite{Ponglertsakul-Winstanley2016}.

These studies are motivated partially by suggestions of massive spin-1 particles as a dark matter ingredient \cite{Holdom1985,ArkaniHamedEtAl2008,Pospelov-Ritz2008,GoodsellEtAl2009}  and partially from sheer curiosity as for the new structures with new properties that may be composed by massive vector particles.

All this effort has been concentrated in spherically symmetric solutions. However, the analogous cylindrically-symmetric solutions are interesting and relevant as well.
Cylindrically-symmetric localized solutions of a self-interacting complex scalar field with a conserved global charge -- so called Q-tubes --
have been constructed and studied in \cite{Sakai:2010ny,Tamaki:2012yk,Volkov-Wohnert2002} and extended to gravity in \cite{Brihaye:2013ita}.

This paper is devoted to the study  of solutions which we name Proca Tubes or Proca Q-Tubes. We will find that as for the spherically-symmetric solutions, gravity is indispensable for localized cylindrically-symmetric solutions in the pure Proca system, while the additional terms in the potential (\ref{potential}) are enough to guarantee localized solutions even in flat space. However, taking gravity into account introduces significant changes with respect to the spherically-symmetric case, such as a much smaller region in parameter space where solutions exist. Moreover, this kind of solutions are not asymptotically conic (which is the analog of asymptotic flatness in our case), but are of the Kasner type.

\section{The Model, Ansatz and Field Equations}\label{The Model}
\setcounter{equation}{0}

The field equations for the self-interacting vector field as derived from (\ref{matterLag}) are
\be \label{ProcaEq}
\nabla_\mu F^{\mu\nu} - 2\frac{dU}{d\psi} A^\nu = 0
\ee
which are supplemented with the constraint (the analog of the Lorentz condition for the Maxwell field)
\be \label{ProcaConstraint}
\nabla_\mu \left( \frac{dU}{d\psi} A^\mu \right)= 0
\ee
The conserved global $U(1)$ current is
\be \label{ProcaCurrent}
J^\mu = -\frac{i}{2}\left( F^{*\mu\nu}A_\nu - F^{\mu\nu}A^*_\nu \right)
\ee

If gravity is ``switched on'', one should solve these equations in a self-consistent way with the Einstein equations which in terms of the Einstein tensor $G_{\mu\nu}$ read:
\be \label{EinsteinEqG}
G_{\mu\nu} + 8\pi G T_{\mu\nu}= 0
\ee
where the energy-momentum tensor is given by
\be \label{energy-mom-tensor}
T_{\mu\nu} = \frac{1}{2}\left(F^{*\lambda}_{\mu}F_{\lambda\nu} + F^{*\lambda}_{\nu}F_{\lambda\mu}\right) + \frac{1}{4} F^*_{\kappa\lambda}F^{\kappa\lambda}g_{\mu\nu} - \frac{dU}{d\psi}\left(A^*_\mu A_\nu + A^*_\nu A_\mu  \right) + U(\psi) g_{\mu\nu}
\ee

As mentioned above, we intend to fill here the gap of cylindrically-symmetric solutions in this system. We assume therefore an appropriate form of metric given by the line element
\be \label{cyl-metric}
ds^2=g_{\mu\nu}dx^\mu dx^\nu= N^2(r) dt^2-H^2(r) dr^2-L^2(r) d\varphi^2-K^2(r) dz^2
,\ee
while for the vector field we assume the radial ``electric'' configuration
\be \label{cyl-VectorField}
A_\mu dx^\mu = e^{-i\omega t}\left( a_0 (r) dt + i a_1(r) dr  \right)
.\ee
The 2 components $a_0 (r)$ and $a_1 (r)$ are assumed to be real. They satisfy the ``Lorentz'' condition which takes now the form
\be \label{LorentzCyl}
\frac{\omega}{N}\left( m^2 + \lambda\psi + \nu \psi^2 \right)a_0 + \frac{1}{HKL}\left[\frac{NKL}{H} \left( m^2 + \lambda\psi + \nu \psi^2 \right)a_1 \right]'=0
\ee
where $\psi$ takes the form $\psi=a_0^2 /N^2-a_1^2 /H^2$.

The field equations (\ref{ProcaEq}) become
\begin{eqnarray}\label{ProcaEqCyl-0}
 \frac{\omega}{N^2}\left( a'_0 - \omega a_1 \right) + \left( m^2 + \lambda\psi + \nu \psi^2 \right)a_1 =0 \\\label{ProcaEqCyl-1}
 \frac{N}{HLK}\left[\frac{KL}{HN} \left( a'_0 - \omega a_1 \right) \right]' - \left( m^2 + \lambda\psi + \nu \psi^2 \right)a_0 =0
\end{eqnarray}
and it is easy to see that substituting (\ref{ProcaEqCyl-0}) into (\ref{ProcaEqCyl-1}) yields (\ref{LorentzCyl}). Alternatively, Eq. (\ref{ProcaEqCyl-0}) is a linear combination of (\ref{LorentzCyl}) and (\ref{ProcaEqCyl-1}).

An important characteristic of the solutions is the global $U(1)$ charge, or in the present situation, the charge per unit length, which may be interpreted here as the number of vector particles per unit length. It is readily obtained from the time component of the conserved current (\ref{ProcaCurrent}):
\be \label{ProcaCharge}
Q= -2 \pi \int_0^\infty dr \frac{KL}{HN} \left( a'_0 - \omega a_1 \right)a_1 =
\frac{2 \pi }{\omega} \int_0^\infty dr \frac{NKL}{H} \left( m^2 + \lambda\psi + \nu \psi^2 \right)a_1^2
\ee
where the second expression was obtained by using (\ref{ProcaEqCyl-0}). We assume of course that the integral converges for the localized solutions we are after.  Without loss of generality we will take $\omega>0$, so $Q>0$ as long as $\lambda^2<4\nu m^2$ holds (see below).

By integrating (\ref{LorentzCyl}) or (\ref{ProcaEqCyl-1}) one obtains the following constraint which is useful as a check on the solutions we will find:
\be \label{ProcaInegrConstraint}
\int_0^\infty dr \frac{HKL}{N}  \left( m^2 + \lambda\psi + \nu \psi^2 \right)a_0 =0
\ee

A second important characteristic is the mass per unit length. Thus we calculate the components of $T_\mu^\nu $ for the cylindrical case which turns out to be diagonal. We find:
\begin{eqnarray}\label{Tmunu}
T_0^0=\frac{\left( a'_0 - \omega a_1 \right)^2}{2N^2 H^2} +\left(m^2 + \lambda\psi + \nu \psi^2 \right)\frac{a_0^2}{N^2} -\left(\frac{m^2}{2} \psi+ \frac{\lambda}{4} \psi^2 + \frac{\nu}{6} \psi^3\right) \\
T_r^r=\frac{\left( a'_0 - \omega a_1 \right)^2}{2N^2 H^2} -\left(m^2 + \lambda\psi + \nu \psi^2 \right)\frac{a_1^2}{H^2} -\left(\frac{m^2}{2} \psi+ \frac{\lambda}{4} \psi^2 + \frac{\nu}{6} \psi^3\right) \\
T_\varphi^\varphi=T_z^z=-\frac{\left( a'_0 - \omega a_1 \right)^2}{2N^2 H^2} -\left(\frac{m^2}{2} \psi+ \frac{\lambda}{4} \psi^2 + \frac{\nu}{6} \psi^3\right)
\end{eqnarray}
So the mass per unit coordinate length is expressed in terms of the mass density $T_0^0$ as
\be \label{InertialMass}
M= 2 \pi \int_0^\infty dr NLHK T_0^0
\ee
Note that the contribution from the potential term to the mass density (i.e. the two last terms of $T_0^0$) is not always positive definite, but it is so for $\lambda^2<4\nu m^2$. This was shown for flat space by Loginov\cite{Loginov2015}, but the same reasoning goes over to curved space. Incidentally, by using this bound on $\lambda^2$ in (\ref{ProcaInegrConstraint}) we can deduce that $a_0 (r)$ of a localized solution must change sign at least once\cite{Loginov2015}, i.e. must have at least one node at a finite value of $r$. The masses themselves that we calculate turn out to be  positive also outside this parameter range. A further significance of the mass and the global charge together is in the stability ratio $M/Qm$ which implies stable solutions when $M/Qm<1$. This will be explained in more details in sec. \ref{PhysProp}.

 Since we prefer to solve Einstein equations in the ``Ricci form''
\be \label{EinsteinEqR}
R_{\mu\nu} + 8\pi G S_{\mu\nu}= 0 \,\,\,\,\,\, , \,\,\, S_{\mu\nu} = T_{\mu\nu} -\frac{T_\lambda^\lambda}{2} g_{\mu\nu}
\ee
we give below the components of $S_\mu^\nu $ for the right-hand side:
\begin{eqnarray}\label{Smunu}
S_0^0=\frac{\left( a'_0 - \omega a_1 \right)^2}{2N^2 H^2} + \left(m^2 + \lambda\psi + \nu \psi^2 \right)\frac{a_0^2}{N^2} -\left( \frac{\lambda}{4} \psi^2 + \frac{\nu}{3} \psi^3\right) \\
S_r^r=\frac{\left( a'_0 - \omega a_1 \right)^2}{2N^2 H^2} -\left(m^2 + \lambda\psi + \nu \psi^2 \right)\frac{a_1^2}{H^2} -\left( \frac{\lambda}{4} \psi^2 + \frac{\nu}{3} \psi^3\right) \\
S_\varphi^\varphi=S_z^z=-\frac{\left( a'_0 - \omega a_1 \right)^2}{2N^2 H^2} -\left( \frac{\lambda}{4} \psi^2 + \frac{\nu}{3} \psi^3\right)
\end{eqnarray}
and the components of Ricci tensor for the left-hand side:
\begin{eqnarray}\label{RicciComp} \nonumber
R_0^0= -\frac{\left(KLN'/H\right)'}{KLNH} \,\,\,\,\,\, , \,\,\, R_\varphi^\varphi = -\frac{\left(NKL'/H\right)'}{KLNH}\,\,\,\,\,\, , \,\,\,R_z^z = -\frac{\left(LNK'/H\right)'}{KLNH}\\
R_r^r = -\frac{1}{H^2}\left(\frac{N''}{N}+\frac{L''}{L}+\frac{K''}{K}
- \frac{H'}{H}\left(\frac{N'}{N}+\frac{L'}{L} + \frac{K'}{K}\right) \right)
\end{eqnarray}
We still have a gauge freedom reflected in the metric component $H(r)$ which we choose from now on to be $H(r)=1$, so we get 3 second order equations for the 3 remaining metric components. By writing the field equations in a dimensionless form, it becomes obvious that there are only 2 independent free parameters in this system. So we replace $r$ by $x=mr$ and scale the components $a_0$ and $a_1$ by a mass scale $\mu=m^{1/2}/\nu^{1/4}$. A dimensionless gravitational coupling constant $\gamma=8\pi G \mu^2=8\pi G m/\nu^{1/2} $ appears naturally.
This way we get
\begin{eqnarray}\label{EinsteinCyl0}
\frac{\left(KLN'\right)'}{KLN} -\gamma \left[  \frac{\left( a'_0 - \omega a_1 \right)^2}{2N^2} + \left(1 + \lambda\psi + \psi^2 \right)\frac{a_0^2}{N^2} -\left( \frac{\lambda}{4} \psi^2 + \frac{1}{3} \psi^3\right) \right]=0\\ \label{EinsteinCylphi}
\frac{\left(NKL'\right)'}{KLN} + \gamma \left[  \frac{\left( a'_0 - \omega a_1 \right)^2}{2N^2} + \left( \frac{\lambda}{4} \psi^2 + \frac{1}{3} \psi^3\right)\right]=0\\ \label{EinsteinCylz}
\frac{\left(LNK'\right)'}{KLN}+ \gamma \left[  \frac{\left( a'_0 - \omega a_1 \right)^2}{2N^2} + \left( \frac{\lambda}{4} \psi^2 + \frac{1}{3} \psi^3\right)\right]=0
\end{eqnarray}
All quantities in these equations are dimensionless. In particular $\omega$ stands for $\omega/m$ and $\lambda$ here is actually $\lambda \mu^2/m^2=\lambda/(m\nu^{1/2})$ in terms of the original parameters. A fourth equation which is not independent but still useful as a check is the $(rr)$ component of (\ref{EinsteinEqG}) which is first order:
\be \label{EinsteinCylGrr}
\frac{N'}{N}\frac{K'}{K}+\frac{K'}{K}\frac{L'}{L} + \frac{L'}{L}\frac{N'}{N}+
\gamma\left[\frac{\left( a'_0 - \omega a_1 \right)^2}{2N^2 H^2} -\left(m^2 + \lambda\psi + \nu \psi^2 \right)\frac{a_1^2}{H^2} -\left(\frac{m^2}{2} \psi+ \frac{\lambda}{4} \psi^2 + \frac{\nu}{6} \psi^3\right)\right]=0
\ee

The dimensionless version of the vector field equations (\ref{LorentzCyl})-(\ref{ProcaEqCyl-1}) are obtained formally by taking $m=1$ and $\nu=1$ and replacing $r$ by the dimensionless radial variable $x$. And of course using the same coordinate condition $H(x)=1$. We will not write them again here. We also comment that the dimensionless version of the charge and mass are obtained by the same way exactly. They are related to one another by $\bar{M}=M/\mu^2$ and $\bar{Q}=Qm/\mu^2$. The stability ratio may be calculated both ways: $M/Qm = \bar{M}/\bar{Q}$.

\section{Proca Q-tubes in flat space} \label{Proca Q-tubes}
\subsection{Boundary Conditions}
\setcounter{equation}{0}

Let us first discuss the probe limit of the equations, i.e. the case $\gamma = 0$.
In this case, the relevant equations reduce to a system for the fields $a_0$ and $a_1$
which can also be derived directly from the following effective Lagrangian:
\be
 L_{eff} = x\left(- \frac{1}{2} (a_0')^2 + \omega a_0' a_1 - V_{eff}(a_0,a_1)\right) \ \ , \ \
V_{eff}(a_0,a_1) = \frac{\omega^2}{2} a_1^2
+ (\frac{1}{2} \psi + \frac{\lambda}{4} \psi^2 + \frac{1}{6} \psi^3 ) \ \ , \ \ \psi = a_0^2 - a_1^2
\ee

In order to formulate a boundary value problem on the interval $[0, \infty)$,
we found it convenient to use a form where
the equations for $a_1$ and $a_0$   are respectively of the first and second order.
Three conditions on the boundary  have then to be imposed.
The regularity of the solutions on the axis of symmetry $x=0$ imposes $a_1(0)=0$ and $a'_0(0)=0$.
These conditions are completed by demanding $a_0(\infty)=0$.

Inspecting the possible asymptotic behavior of the fields, we found two possibilities. The first is:
\be
\label{type_I}
       a_0 (x)    \propto \frac{e^{- \sqrt{1-\omega^2} \hspace{0.075cm} x}}{\sqrt x} \ \ , \ \
       a_1 (x)= - \frac{\omega}{1 - \omega^2} \frac{d a_0}{dx}
\ee
The solutions of this type will be referred to  as Type-1.  The other possibility is the following:
\be
\label{type_0}
         a_0(x) = - \frac{A}{\omega x} + o(1/x^2) \ \ , \ \
          a_1(x) = A  - \frac{\omega}{1 - \omega^2} \frac{d a_0}{dx}
\ee
where the constants $A,\omega$ are related by $\omega^2 +  \lambda A^2 - \nu A^4=1$. We will refer
to these solutions as being Type-0. We note that the possibility (\ref{type_0}) is specific
to the vector field system and has no counterpart for the scalar case.

The Type-0 boundary conditions comes from the fact that the effective potential $V_{eff}$
admits some extra critical points, apart from the origin $a_0=a_1=0$. In particular,
(as observed in \cite{{Loginov2015}}) if
 \be
 \label{omega_min}
 \omega^2 > \omega_m^2 \equiv 1 - 3 \lambda^2/16 \ \ ,
 \ee
one of the critical points occurring leads to a negative value of $V_{eff}$.
As a consequence, solutions of Type-0 can be expected for appropriate values of $\lambda$ and $\omega$.

\begin{figure}[b!]
\begin{center}
{\includegraphics[width=8cm]{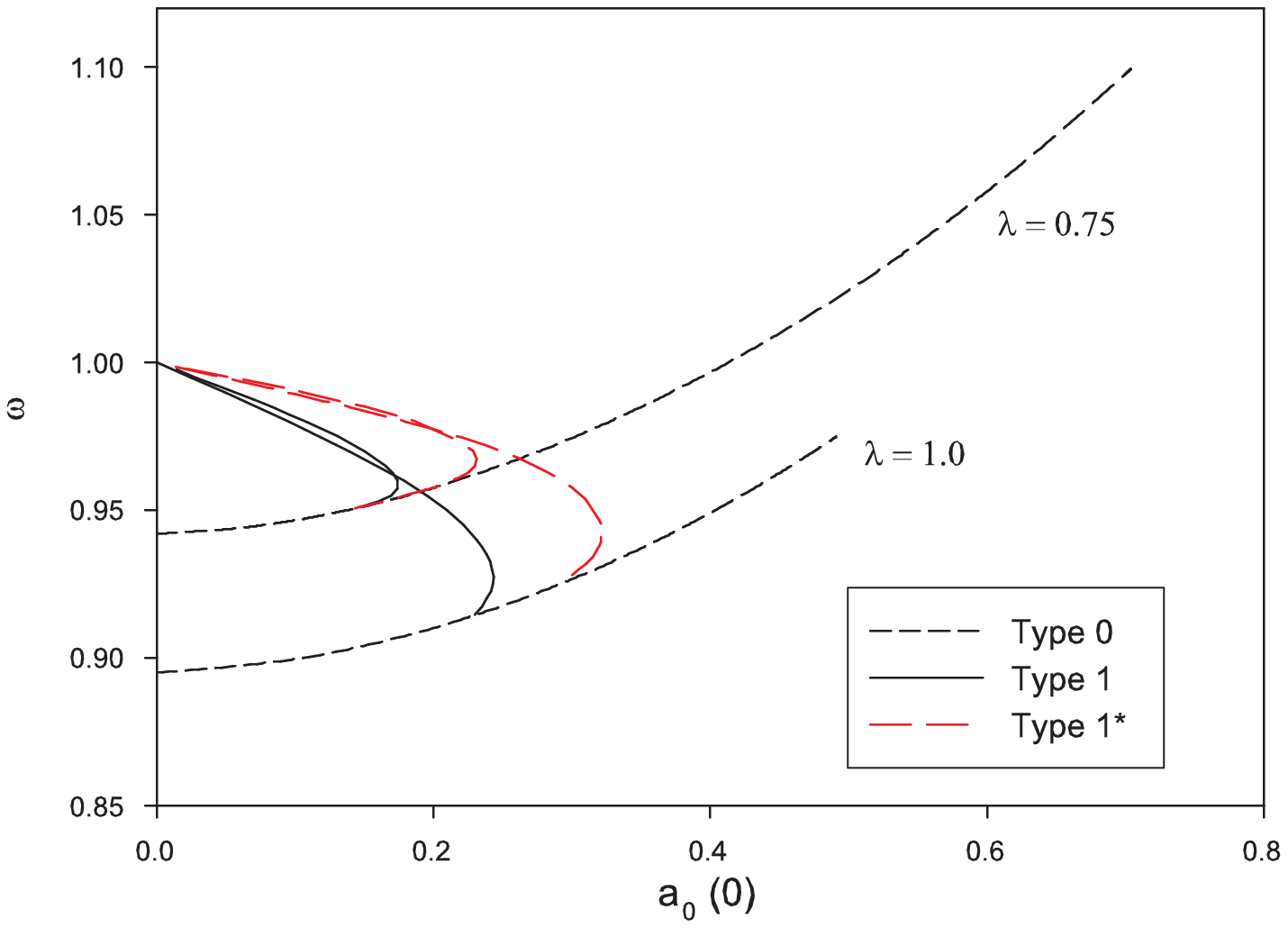}}
{\includegraphics[width=8cm]{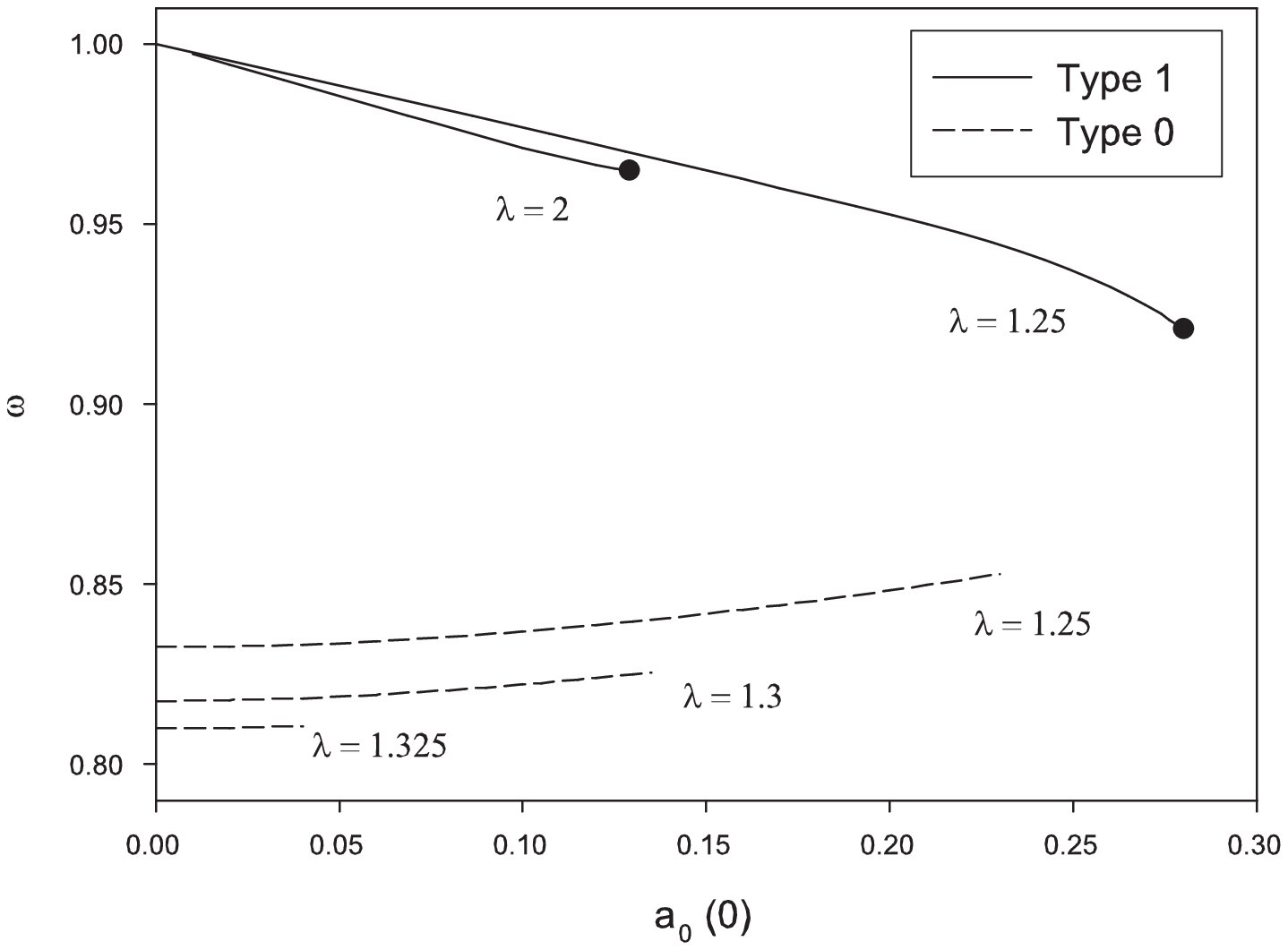}}
\caption{Left: Frequency $\omega$ versus  $a_0(0)$
for  $\lambda= 0.75$ and $\lambda =1.0$ for the Type-0
(black-dashed),  Type-1 (black-solid) and Type-1* (red-dashed) solutions. Recall that all quantities are dimensionless.
Right: $\omega$ versus  $a_0(0)$ of Type-0 and Type-1 solutions for higher values of $\lambda$ - strong coupling case explained in sec. \ref{SC}.
\label{type_0_I}
}
\end{center}
\end{figure}

Once the boundary conditions are imposed, and with a given choice
of the self-interacting potential (actually, by choosing the rescaled parameter $\lambda$ only),
the system can be solved for continuous values of the frequency $\omega$ or
of the central value $a_0(0)$;
these quantities are related through the equations - see e.g. Fig. \ref{type_0_I}.
The set of solutions labeled by $a_0(0)$ or $\omega$ then constitutes a branch of Proca Q-tubes. The main challenge
is to determine the pattern of these solutions for different choices of the coupling constant $\lambda$.

Since the two asymptotic forms (\ref{type_I}) and (\ref{type_0})
are consistent with the  boundary conditions, the  corresponding  solutions have been emphasized.
These families present quite different properties, in particular:
\begin{itemize}
\item {\bf Type-0 :} These solutions have
 $a_1(\infty) \neq 0$ and $a'_1(\infty) = 0$.
   They have no finite mass
per unit length as seen from Eqs. (\ref{Tmunu}), (\ref{InertialMass}).
Nevertheless they turn out useful to understand of the pattern of solutions.
\item {\bf Type-1 :} These solutions  have
 $a_1(\infty) = 0$ and $a'_1(\infty) = 0$. These solutions have a finite mass
per unit length. These are Proca Q-tubes.
\end{itemize}
The occurrence (and eventually the co-existence) of
Type-0 and Type-1 solutions depends, as we will see, on
the potential (i.e. the parameter $\lambda$) and on the value of  $a_0(0)$.

Let us finally point out that Type-1 solutions split into several classes according their radial excitation number which is in a one-to-one correspondence with the number of nodes of $a_0(x)$ and $a_1(x)$. We will denote these classes as Type 1* (for the first excitation), Type 1**, etc.

\subsection{Numerical results}
\setcounter{equation}{0}
The  Type-1 solutions can be constructed in a way technically inspired
from the construction of spherically symmetric Q-balls in three spatial dimensions.
Since we do not expect closed form solutions to exist, we construct
the solutions by numerical methods. We used in particular the routine COLSYS \cite{colsys}.

First we pick a value for $\lambda$. Obviously, setting $a_0(0)=0$,
the equations are fulfilled by the trivial configuration $a_0(x)=a_1(x)=0$.
Then by increasing gradually the parameter $a_0(0)$, non trivial solutions to the boundary value problem emerge provided the frequency $\omega$ is fine-tuned appropriately.
The numerical results show that  solutions exist for $\omega_c < \omega < \omega_M$
where $\omega_M$ can be determined by
using an argument presented in \cite{Volkov-Wohnert2002}. Namely $\omega_M = -2 \frac{d U}{d \psi}(0)$
which, in our scaling, corresponds to $\omega_M = 1$ (the issue of $\omega_c$ will be discussed next).
The limiting value $\omega \to \omega_M$ corresponds to $a_0(0) \to 0$.
The $\omega - a_0(0)$ relations for the solutions
are represented by the solid lines in Fig. \ref{type_0_I} for several values of $\lambda$.
The next task is to determine the value $\omega_c$; in fact
the critical phenomenon  limiting  the solutions for $\omega \to \omega_c$
depends  on the value of the constant $\lambda$ as we now discuss.


\subsubsection{Weak coupling:} \label{WC}
\begin{figure}[t]
\begin{center}
{\includegraphics[width=9cm]{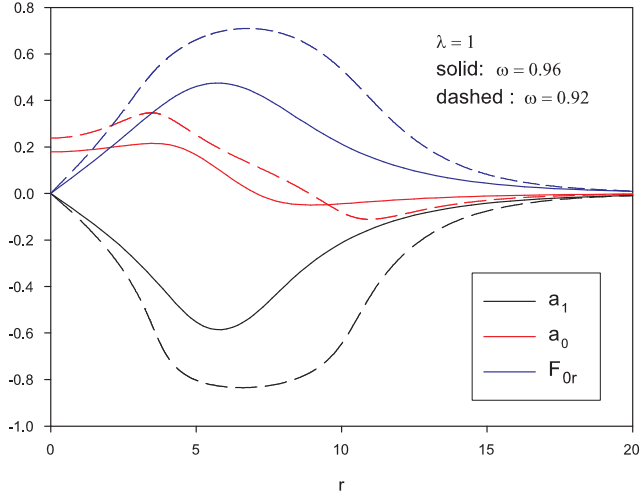}}
\end{center}
\caption{Profiles of Type-1 solution with  $\lambda = 1$; the solid (resp. dashed) lines
correspond to $\omega = 0.96$ and $a_0(0) = 0.1787$ (resp. $\omega = 0.92$ and $a_0(0) = 0.238$).
\label{type_I_cr}
}
\end{figure}

Let us first examine the case of small $\lambda$ (typically for $0 < \lambda \leq  1.15$).
First we recall that for type-1 solutions, $a_0 (x)$ has one node at some $x>0$, while $a_1 (x)$ has a node at $x=0$ and a minimum at some $x>0$. While decreasing $\omega$  for Type-1 solutions it turns out that
$a_1 (x)$  has the tendency to stay very close to its minimal value
for a larger and larger  interval of $x$.
This is illustrated in Fig.  \ref{type_I_cr} where we set $\lambda =1$ for
the solutions with  $\omega=0.96$ and $\omega=0.92$ (note: here we find $\omega_c \approx 0.9125$).

The  Type-0  solutions have been constructed as well and, like Type-1,
are characterized by the value $a_0(0)$ and the relation of this parameter with frequency $\omega$.
The $\omega - a_0(0)$ relation of the Type-0  solutions
corresponding to  $\lambda=1.0$ and  $\lambda=0.75$ are shown by the black-dashed lines
on Fig. \ref{type_0_I}.
Interestingly, this figure  reveals that the branch of Type-1 solutions bifurcates from a branch
 of Type-0 solutions at the minimal frequency for Type-1 solutions, $\omega=\omega_c$.
 This property, which was confirmed for several values of $\lambda$, contrasts with the case of Q-balls
 where the central value of the field, say $\phi(0)$ can take arbitrarily large values. We will see later that the maximal value of $a_0(0)$ on the bifurcating Type-1 branch is a border between lower and higher mass solutions, where the higher mass ones are more stable than the lower mass Q-tubes.

\begin{figure}[t!]
\begin{center}
{\includegraphics[width=8cm]{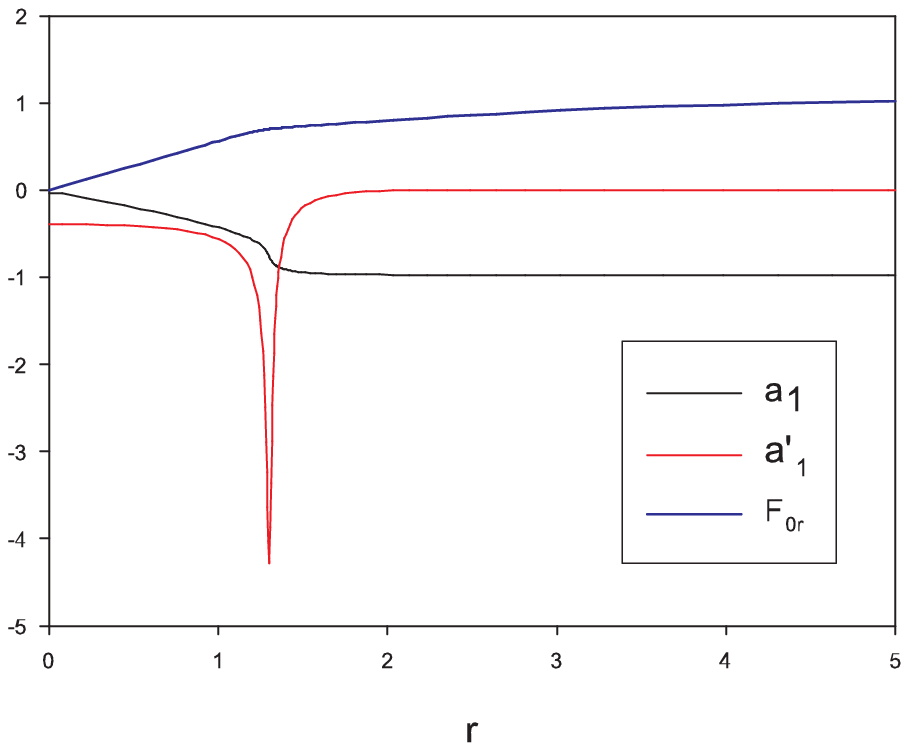}}
{\includegraphics[width=8cm]{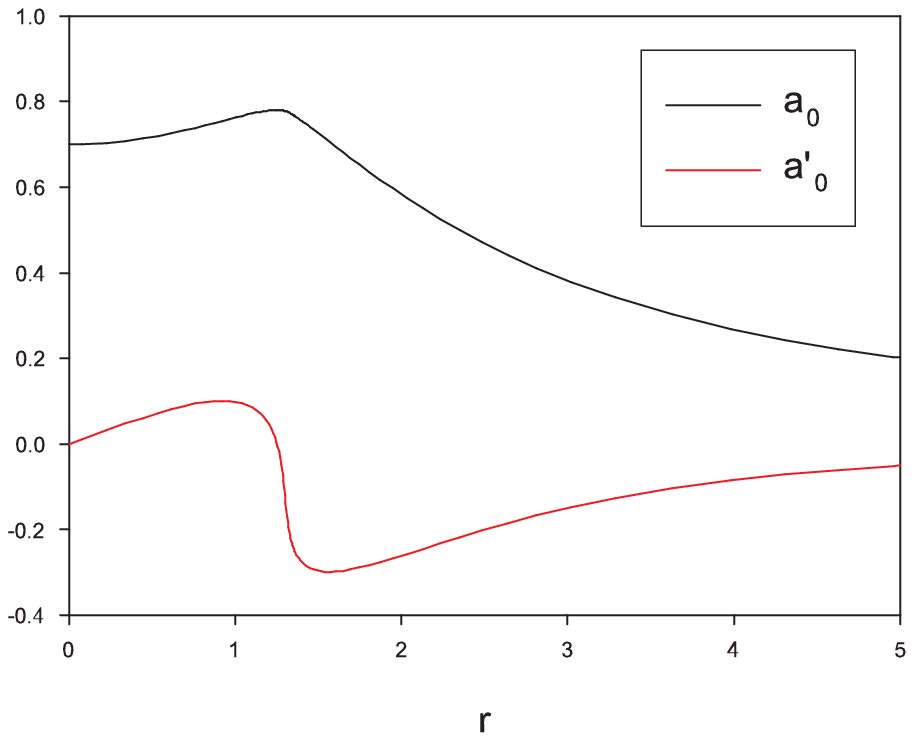}}
\caption{Profiles of a Type-0 limiting solution for $\lambda = 0.75$
with $a_0(0) = 0.70$ (corresponding to $\omega \approx 1.1$). Note that the ``electric field'' $F_{0r}$ is smooth in spite of the shape of $a_0$ and $a_1$.
\label{type_0_cr}
}
\end{center}
\end{figure}

Similarly, the first excited states are found along a curve which branches of the Type-0 curve somewhat higher and so on for the higher excitations as well. The corresponding data is presented by the red-dashed lines
on the $\omega-a_0(0)$ curve  in Fig. \ref{type_0_I}.
\\
\\
Let us summarize a few features
characterizing the Type-0 and Type-1 solutions.
 \begin{itemize}
\item Contrasting with the Type-1, Type-0 solutions  have
             $\lim_{a_0(0) \to 0} \omega = \omega_m(\lambda)$ where $\omega_m(\lambda)$ is determined by Eq. (\ref{omega_min}).
\item While the Type-1 branch approaches the vacuum configuration in the limit $a_0(0) \to 0$,
Type-0 have  non-trivial $a_0 (x)$ and $a_1 (x)$ as a limiting solution.
\item When  $a_0(0)$ approaches a maximal value on the Type - 0 branch, say $a_0 (0)_{max}$,
the corresponding solutions approach a singular configuration
such as illustrated  by Fig. \ref{type_0_cr} for $\lambda = 0.75$
($a_0 (0)_{max} \approx 0.70$, $\omega \approx 1.1$).
\item In this critical limit
the Type-0 field $a_1(x)$  approaches a  wall-shape at a particular
value of the radial coordinate, say $x=x_w$, and  reaches its (non-zero) asymptotic value for $x > x_w$.
As a consequence, the value $|a_1'(x_w)|$ becomes very large.
\end{itemize}

\subsubsection{Strong coupling} \label{SC}
\begin{figure}[b!]
\begin{center}
{\includegraphics[width=8cm]{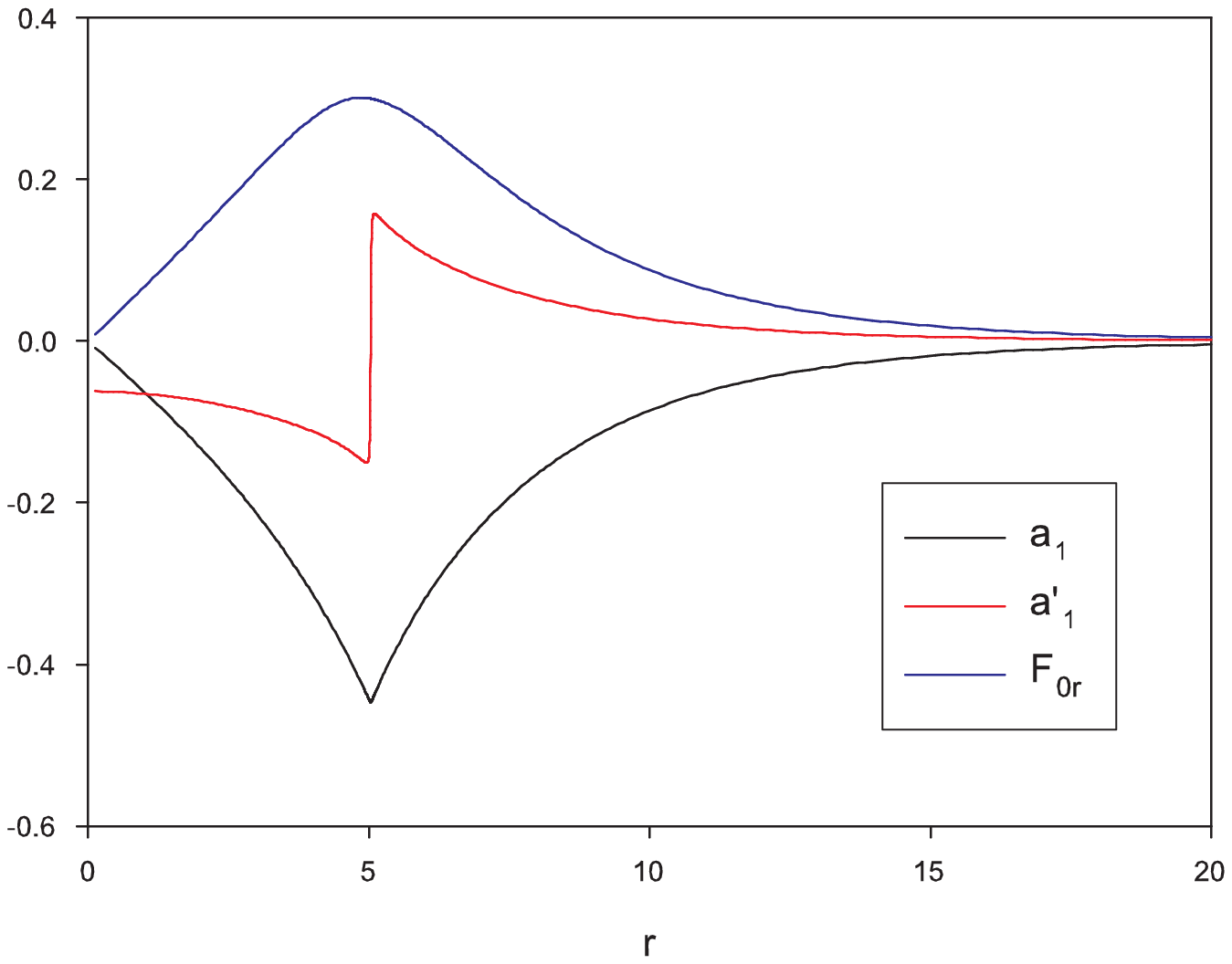}}
{\includegraphics[width=8cm]{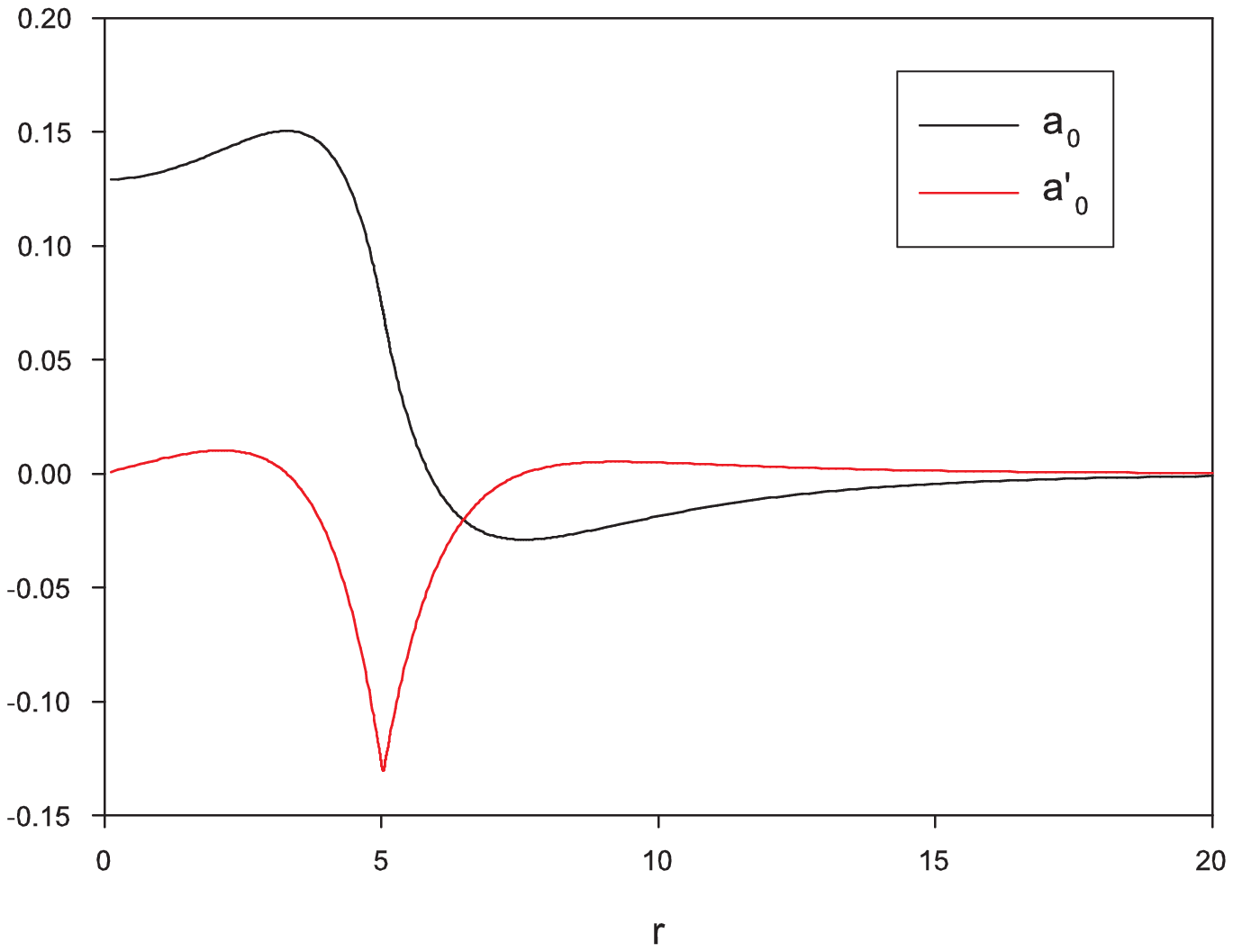}}
\caption{Profiles of a Type-1 limiting solution for $\lambda = 2$ with $a_0(0) = 0.13$ (corresponding to $\omega = 0.965$).
Note that the ``electric field'' $F_{0r}$ is smooth in spite of the shape of $a_0$ and $a_1$.
\label{type_I_cr_bis}
}
\end{center}
\end{figure}

Solutions of both types (Type-0 and Type-1) can be constructed for large values of $\lambda$ (typically $\lambda > 1.15$)
but the branch of Type-1 solutions  ceases to bifurcate into the Type-0 one,
{as shown on the right side of Fig. \ref{type_0_I}.
The numerical results suggest the Proca Q-tubes terminate into a singular (i.e. non-analytical) configuration when a maximal
value of the parameter $a_0(0)$ is approached (this is symbolized  by the bullets on Fig. \ref{type_0_I}.} At the approach of this maximal value
 the Proca fields $a_0(x)$ and $a_1(x)$ behave  differently in the two regions separated by the cylinder $x=x_w$~:
the function  $a_0(x)$  forms a  'wall' at  $x=x_w$. In the neighborhood of the wall the function $a_1(x)$ presents
a 'V-shape' and the function $a_1'(x)$ presents a discontinuity.
The limiting solution for $\lambda = 2$ is shown in Fig. \ref{type_I_cr_bis};
it has $a_0(0) = 0.13$  and a corresponding frequency $\omega \approx 0.965$. Note that the field strength
$F_{0r} = -(a_0'-\omega a_1 )$ is smooth.

Type-0 solutions have been constructed  for $\lambda >  1$ as well. We found that
 the domain of existence of these solutions decreases strongly when increasing $\lambda$
and vanishes for $\lambda \approx 1.34$ but  we have, so far, no explanation for this feature.
In the limit $a_0(0) \to 0$ the frequency  $\omega_m$ (defined at Eq. (\ref{omega_min})) is approached.

\subsubsection{Physical properties} \label{PhysProp}

\begin{figure}[t!]
\begin{center}
{\includegraphics[width=8cm]{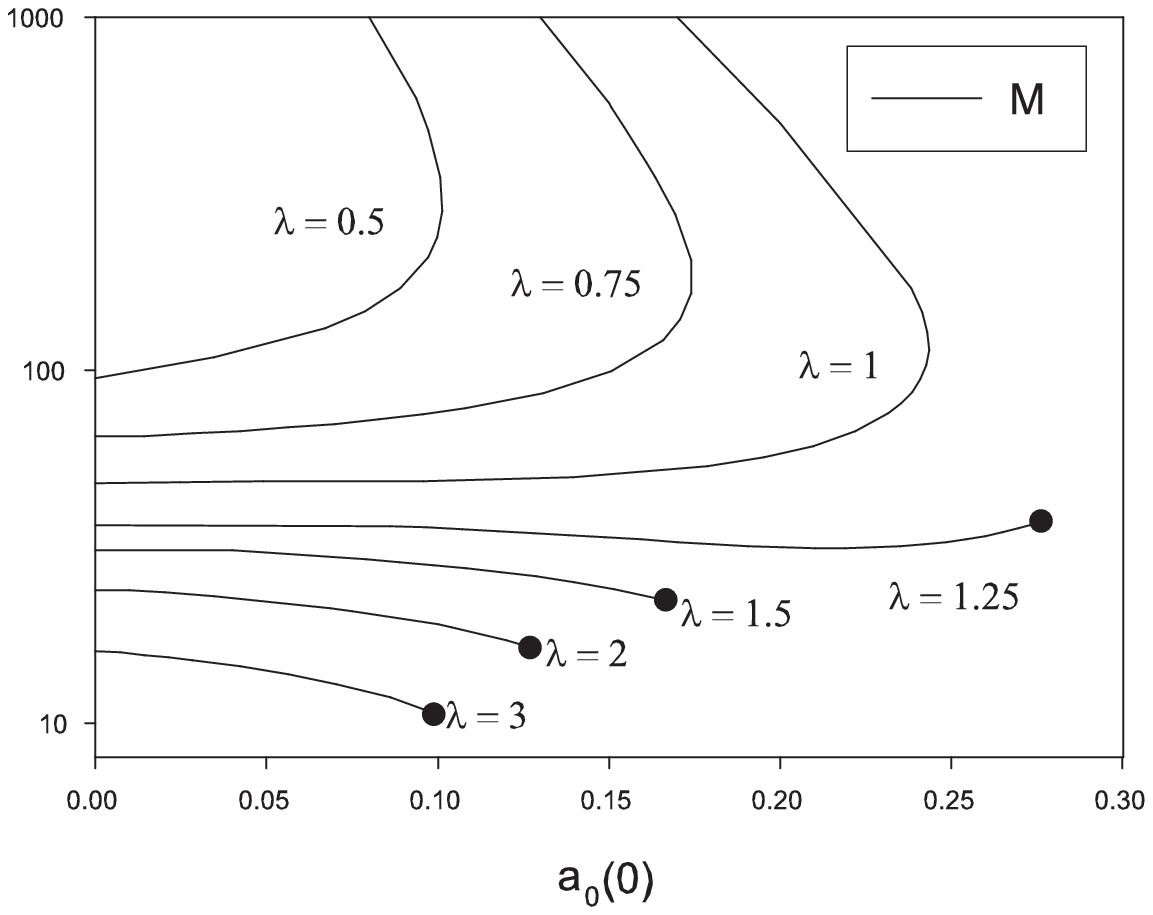}}
{\includegraphics[width=8cm]{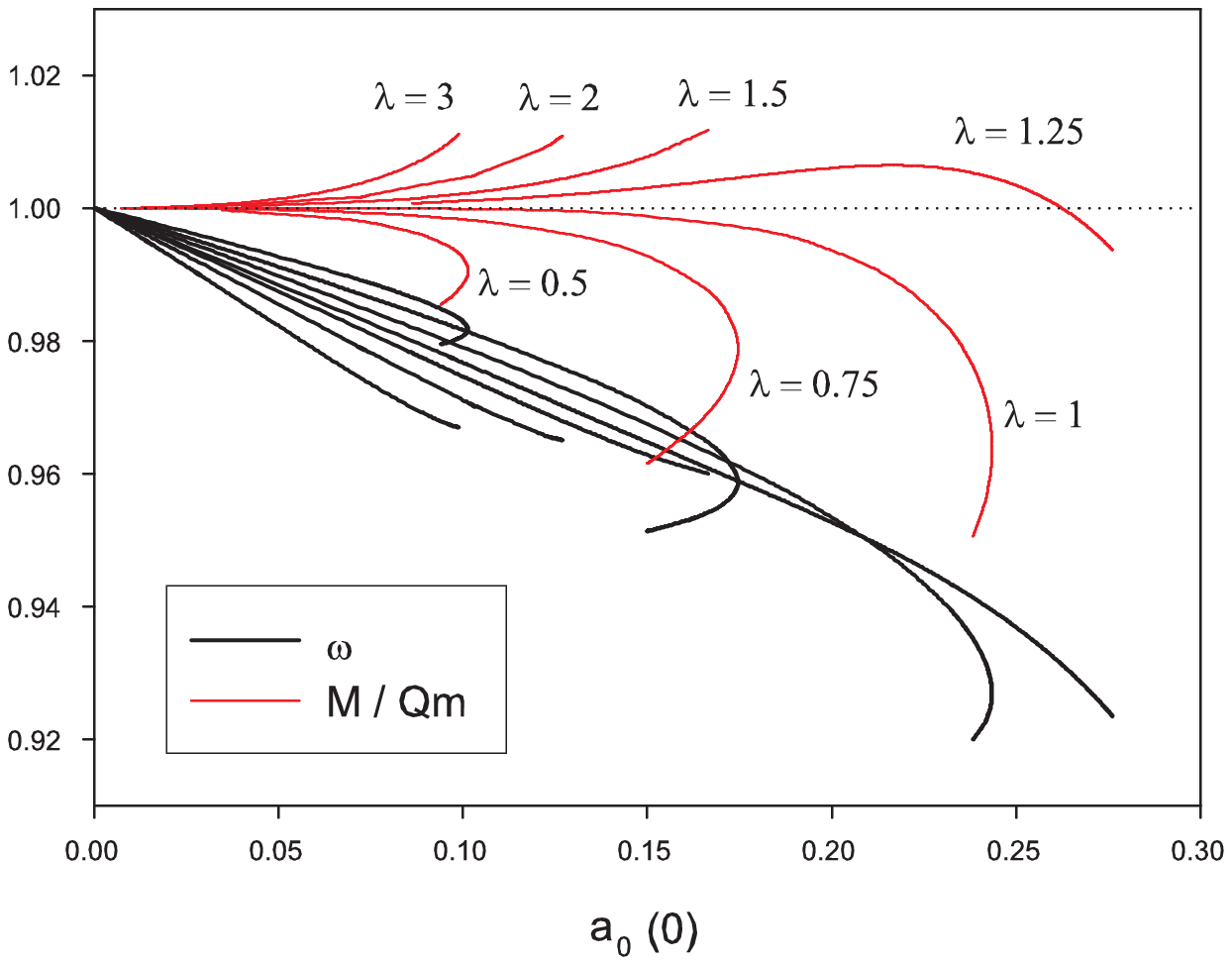}}
\caption{Left: Q-tube mass (logarithmic scale) as a function of $a_0(0)$ for several values of $\lambda$. Right:
The corresponding frequency $\omega$ and the ratio $M/Qm$ as function of $a_0(0)$. Note that the $\lambda$ values of the $\omega$ curves can be inferred from comparison with the higher $M/Qm$ curves.
\label{mass_type_I}
}
\end{center}
\end{figure}

We now present some physical  quantities characterizing the Type-1 (Q-tube) families
namely  the mass $M$ and the charge $Q$.
From these data, the  stability criterium usually used in the study of boson stars (see e.g.  \cite{FriedbergLeeSirlin1976}) can be inferred.
It consists in studying the sign of $B \equiv M/(Qm)-1$; since $Q$ represents the particle number and $m$ the
mass of an individual quantum, $M-Qm$ provides a measure of the classical binding energy of the solution.
 Solutions with $B < 0$ have a positive binding energy per particle and then cannot  decay into $Q$ elementary quanta;
 $B<0$ is usually seen as a necessary condition of stability.

The dependence of the masses of these solutions on the central value $a_0(0)$  is depicted in the left side
of Fig. \ref{mass_type_I} for several values of $\lambda$. For the small values of $\lambda$, the mass diverges as $\omega$ approaches the critical value
in agreement with the fact that these branches approach Type-0 solutions. For large $\lambda$, the
Type-0 and Type-1 branches do not meet and the Type-1 branches terminate in a singular configuration like that of Fig. \ref{type_I_cr_bis}, symbolized by the bullets on the plot.Note also the finite mass (and also charge)  as $a_0 (0)\rightarrow 0$. This is a common feature of Q-balls and Q-tubes \cite{KleihausEtAl2005, BrihayeEtAl2015} and occurs since the fields fall off to zero very slowly such that the integrals for the mass and charge stay finite.

\begin{figure}[b!]
\begin{center}
{\includegraphics[width=8cm]{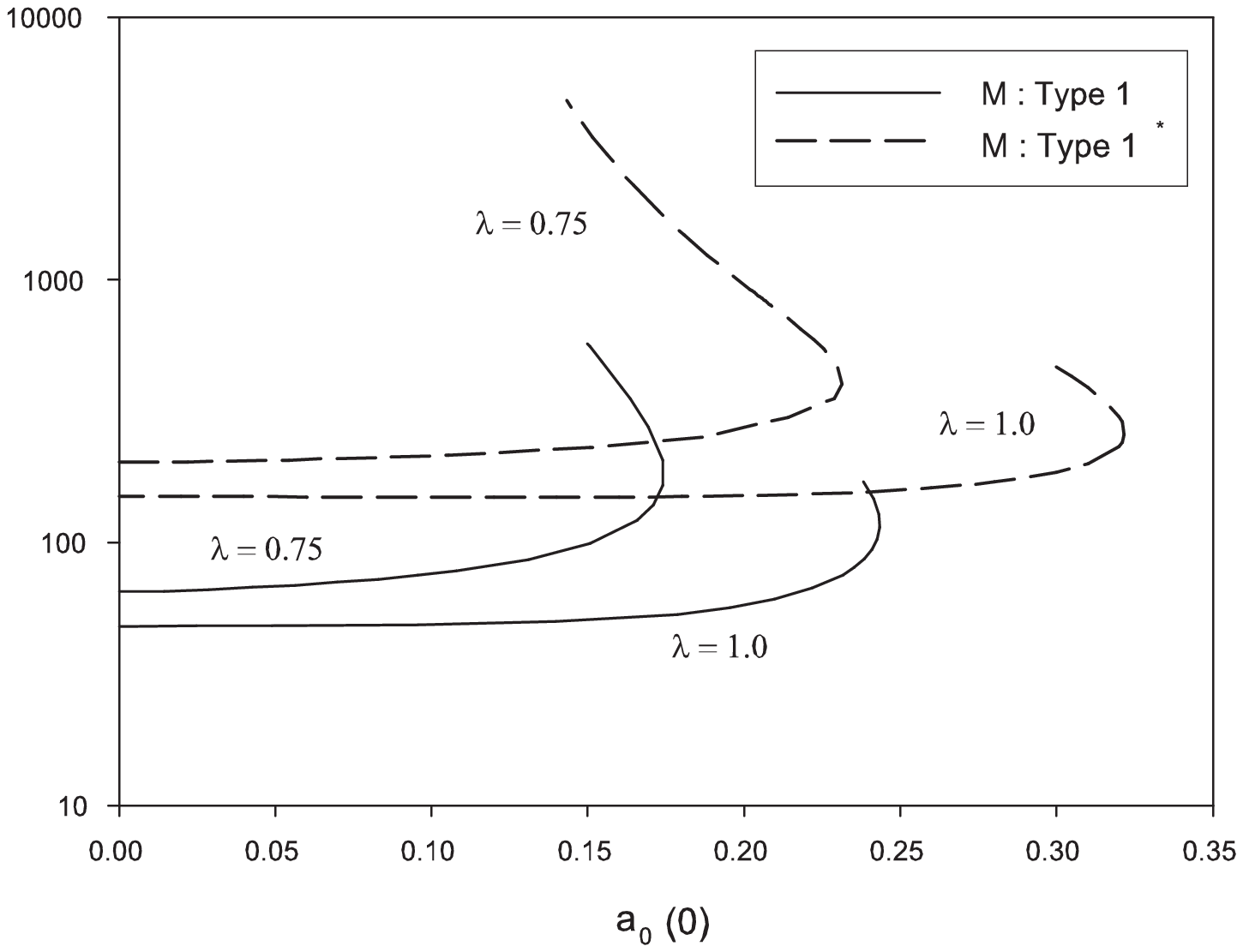}}
{\includegraphics[width=8cm]{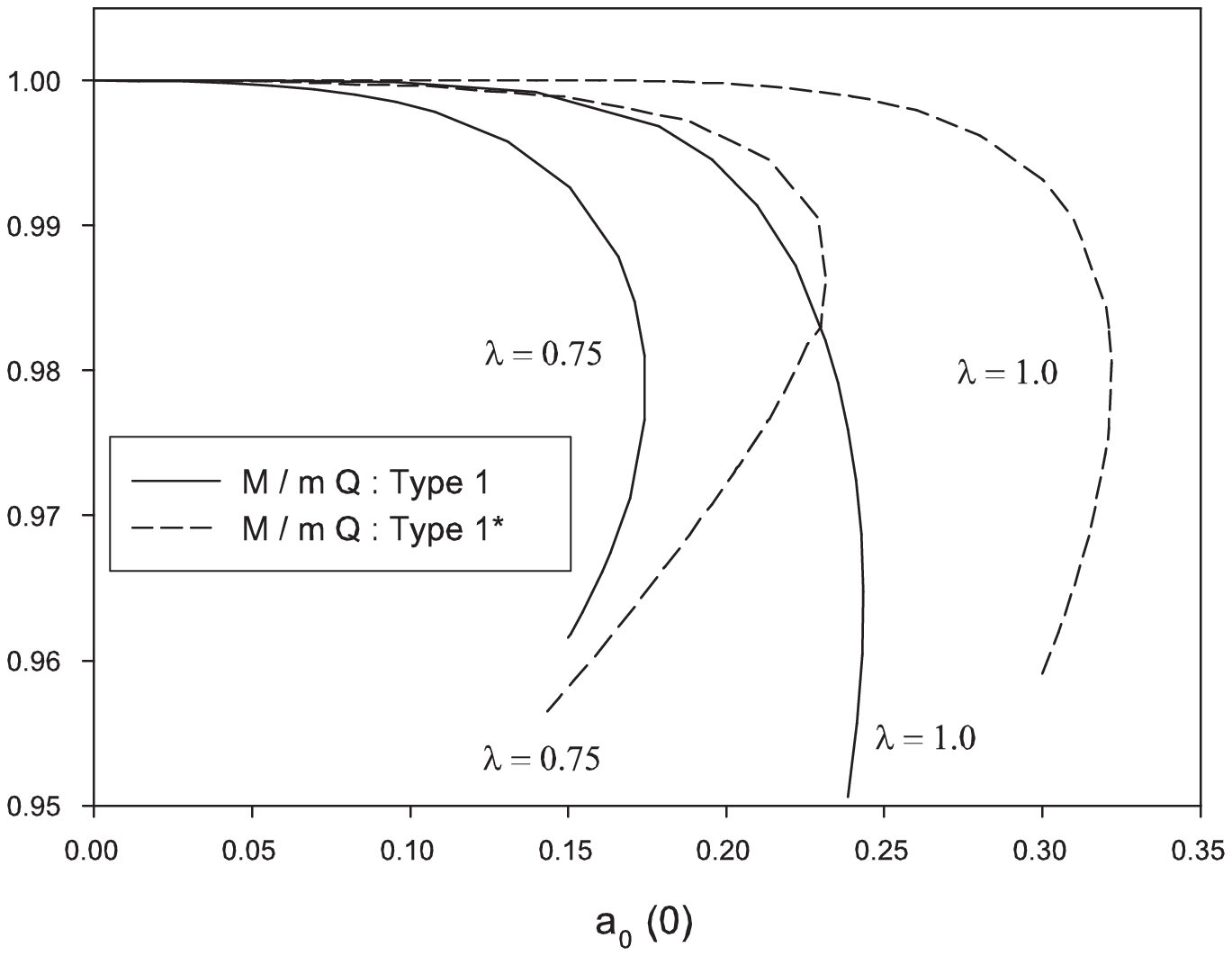}}
\caption{Left:  Mass (logarithmic scale) as a function of $a_0(0)$ for $\lambda=0.75$ and $\lambda=1$ of the Type-1 (solid lines) and Type-1$^*$ (dashed lines)
 Proca Q-tubes. Right:
The correponding  ratio $M/Qm$.
\label{MassesGround+Exctd}
}
\end{center}
\end{figure}

The right panel of the figure shows the $a_0(0)$ dependence of $\omega$
and of the ratio $M/Qm$ discussed above. For the weak coupling solutions, $a_0(0)$ first increases as $\omega$ decreases from 1, reaches a maximum and then decreases while $\omega$ tends to its minimum value of $\omega_c$ at the branching point. The Q-tubes of this lowest part of the curve have the highest mass and are more stable then the higher $\omega$ tubes. We see that stable solutions ($M/Qm < 1$) exist up to about a little higher than $\lambda=1$. At $\lambda=1.25$ most of the Q-tube solutions become unstable, apart from a small interval  of  $a_0(0)$ or $\omega$. For larger values of $\lambda$ (i.e. strong coupling), all solutions are unstable. We also note that the condition for the positivity of $T_0^0$, whose dimensionless form reads $\lambda^2<4$, seems to play no role in this context. There exist positive mass solutions above $\lambda=2$ and unstable solutions below $\lambda=2$.

We also constructed several classes of radially excited states of Proca Q-tubes. Figure \ref{MassesGround+Exctd} presents the $M - a_0(0)$ curves of the ground states and first excited states for two values of $\lambda$. In view of these solutions and since radially excited states in scalar Q-balls are well known to exist \cite{Volkov-Wohnert2002,KleihausEtAl2005,Mai-Schweitzer2012},
 it can be expected that excited Proca Q-balls also exist. However, there have been no reports about it so far.


  \section{Gravitating Proca Q-tubes}\label{GravitQT}
\setcounter{equation}{0}

 In this section, we study the effect of gravity on the Q-tubes discussed in the previous section;
 assuming a minimal coupling to gravity (Eq.(\ref{totalAction}))\footnote{Proca field with non-minimal coupling to gravity was discussed by Fan\cite{Fan2016} in the context of black hole solutions. We intend to give attention to this case in a future publication.}. The equations then involve the Newton constant which, after  an appropriate rescaling,
 appears through the effective gravitational parameter $\gamma \equiv 8 \pi G \mu^2$.
Due to the occurrence of  two independent
coupling constants in the equations ($\gamma$ and $\lambda$), we did not study the solutions in a systematic way but focused on
 a few particular values  which we considered as representative.
 The values chosen for reporting our results ($\gamma = 1/40$, $\gamma= 1/80$ and $\gamma = 1/160$)
  are large but this choice has a double purpose: (i) facilitate  the numerical analysis,
 (ii) enhance the effects of gravity which could  be too small to be appreciated
 with values of $\gamma$ corresponding to ``ordinary'' mass scales of the order of a TeV, or even
 to the GUT scale.
 
 \subsection{Boundary value and asymptotics} \label{BVAsymp}
 As far as Proca equations for matter are concerned, the boundary conditions are identical to the non-gravitating case.
 The Einstein equations imply the following conditions for the metric to be regular on the axis of symmetry:
 \be
        N(0) = 1 \ \ , \ \ N'(0) = 0 \ \ , \ \ L(0)=0 \ \ , \ \ L'(0) = 1 \ \ , \ \ K(0) = 1 \ \ , \ \ K'(0)= 0 \ \ .
 \ee
Perhaps the main feature characterizing the gravitating solutions is the fact that the underlying space-time
 is not asymptotically flat (or conic). Indeed an analysis of the asymptotic form of the field reveals that
 the metric is    asymptotically of the Kasner-Type:
 \be
 \label{kasner_power}
         N(x)|_{x \to \infty} \sim x^a \ \ , \ \ L(x)|_{x \to \infty} \sim x^b \ \ , \ \ K(x)|_{x \to \infty} \sim x^c \ \
         , \ \ a+b+c=a^2+b^2+c^2 = 1 \ \ .
 \ee
Kasner solution is a well-known parametrization of the most general cylindrically-symmetric and static vacuum solution of Einstein equations and may be obtained easily from the vacuum version of Eqs. (\ref{EinsteinCyl0})-(\ref{EinsteinCylGrr}).

The values $a,b,c$ are generic and can be reconstructed from the numerical solutions. They are also related to the internal structure of the source (Proca tubes in our case) and may be expressed in terms of integrals of  $S_\mu^\nu $ over the tube cross section \cite{ColdingEtAl1997}. More detailed study of the asymptotic behavior reveals that the power $a$ must satisfy $a>0$. This constitutes a major difference between the gravitating Proca tubes and the ones of the probe limit.
 
 The reason that these solutions cannot be asymptotically conic, is that this kind of solution must have $N(x)=K(x)$ hence $a=c=0$ (and $b=1$) which is related  to the corresponding ``boost symmetry'' of the source - i.e. the components of $T_\mu^\nu $ or $S_\mu^\nu $ and moreover to the fact that the integrals of $S_0^0 $ and $S_z^z $ over the tube cross section vanish  \cite{ColdingEtAl1997}.
 Incidentally, by looking at (\ref{EinsteinCylphi})-(\ref{EinsteinCylz}) we notice now a symmetry between $L(x)$ and $K(x)$. However, it does not follow in this case that $L(x)=K(x)$, neither $b=c$ , since  it is  inconsistent with the boundary conditions.
 As for the vector field, we find
 \be
 \label{a0Asympt-Grav}
 a_0(x)|_{x \to \infty}  \sim  x^p \exp(-x) \ \ , \ \ \ p = a - 1/2\ \ .
 \ee
  
\begin{figure}[t!]
\begin{center}
{\includegraphics[width=8cm]{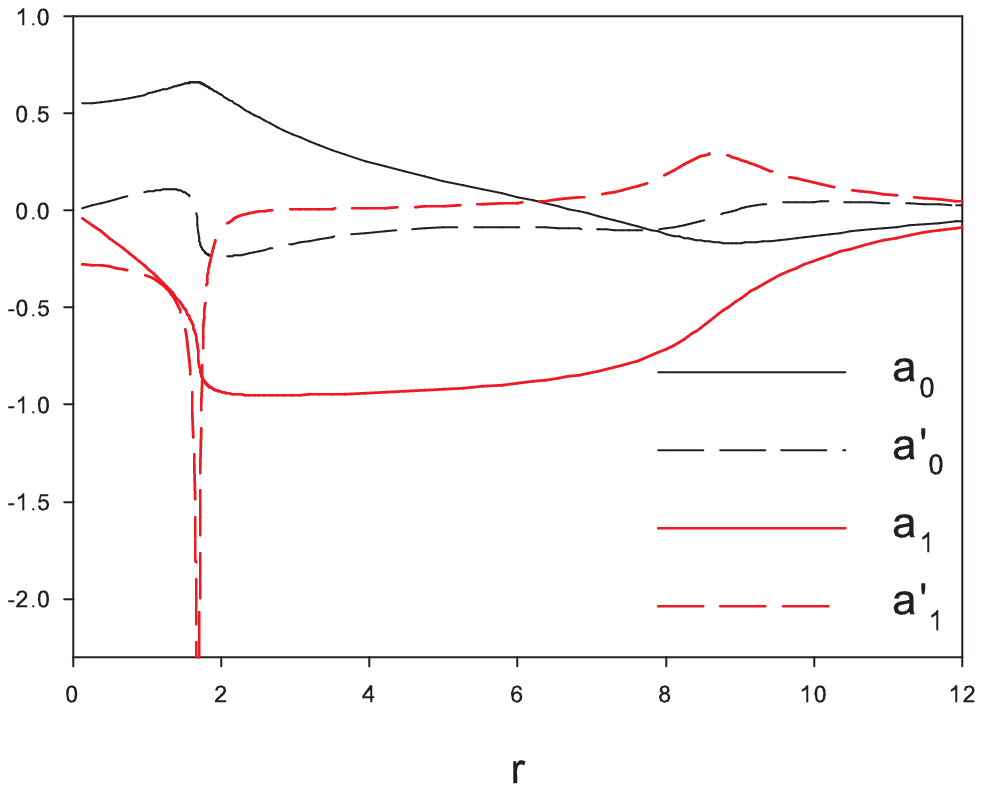}}
{\includegraphics[width=8cm]{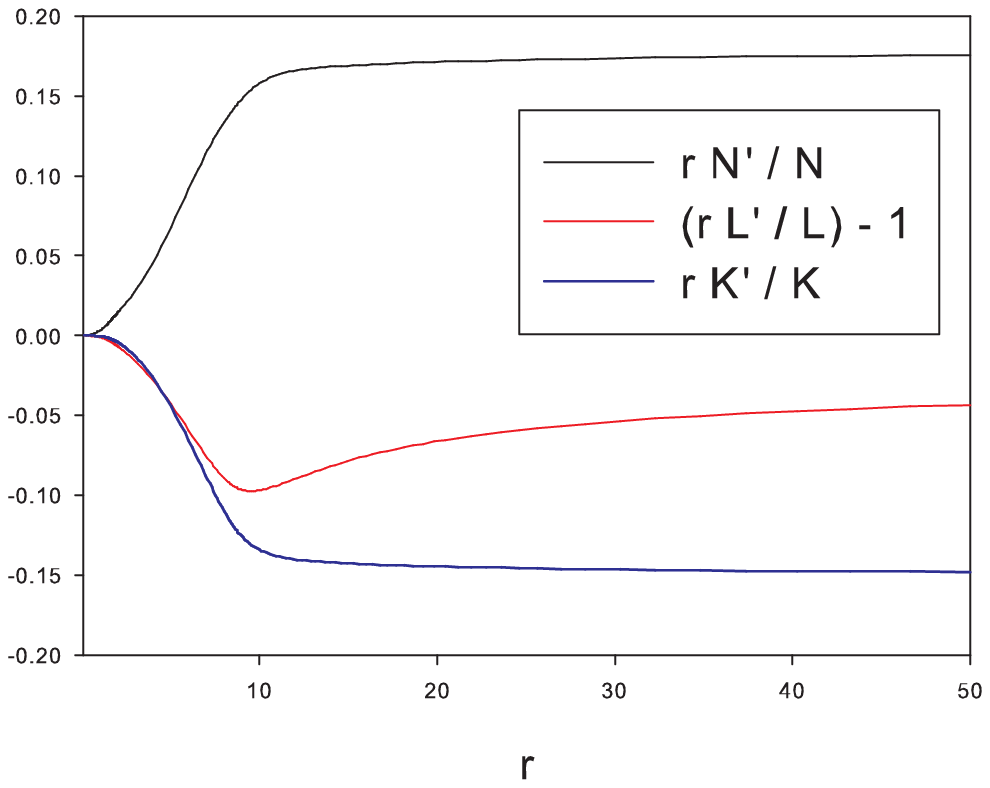}}
\caption{Left: Profiles of the vector field components $a_0 (x)$, $a_1 (x)$ and their derivatives for the
 solution corresponding to  $\lambda=1$, $\gamma = 1/40$ and $a_0(0) = 0.55$.
Right: The corresponding functions   $rN'/N$, $(rL'/L)-1$ and $rK'/K$. This way of presenting the metric components is aimed to stress the asymptotic power-law behavior.
\label{grav_profile}
}
\end{center}
\end{figure}

\begin{figure}[b!]
\begin{center}
{\includegraphics[width=8cm]{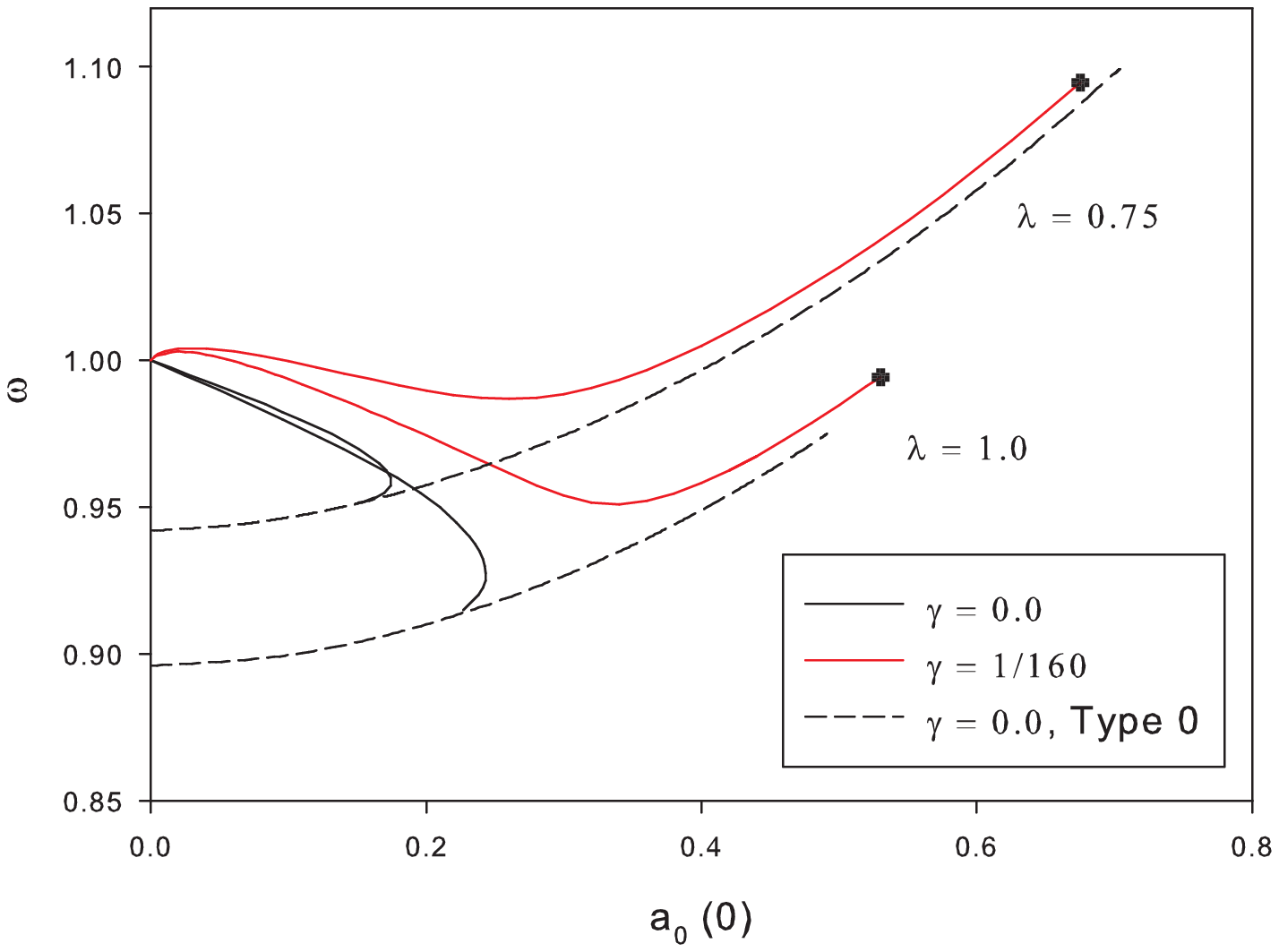}}
{\includegraphics[width=8cm]{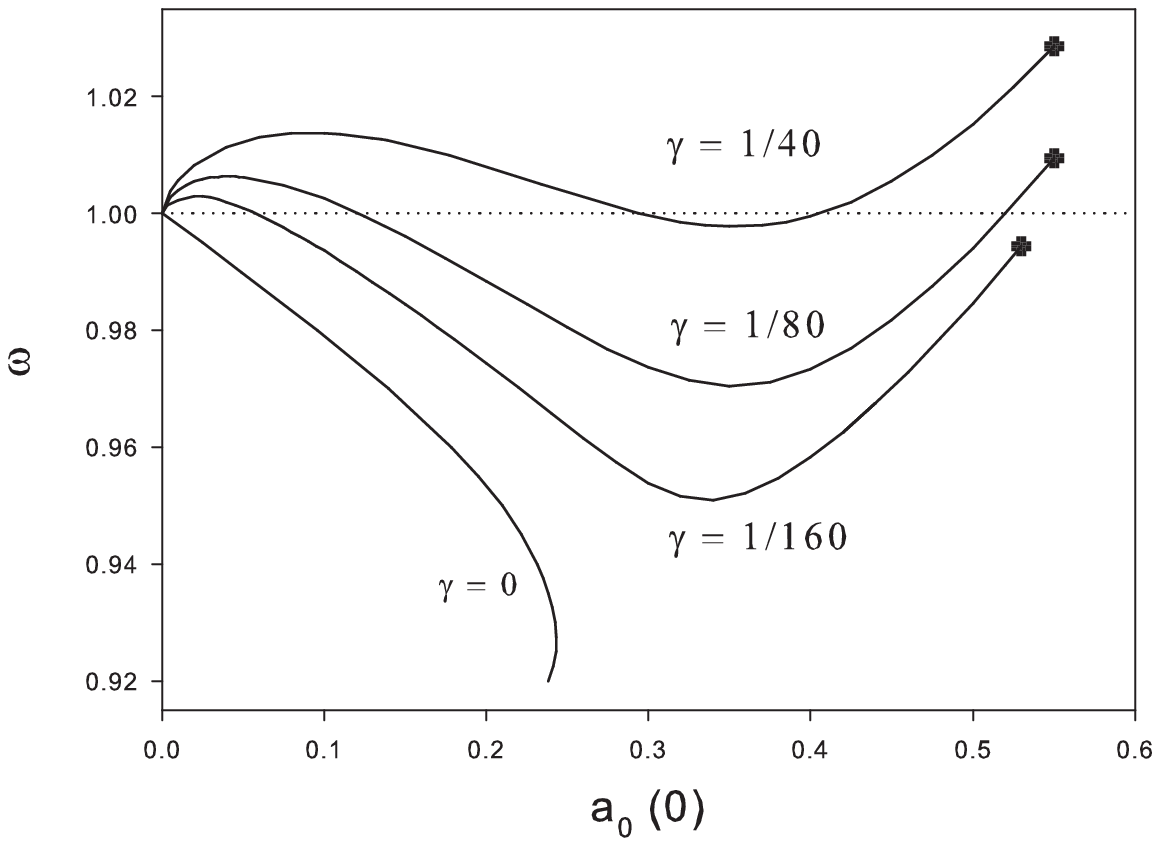}}
\caption{The effect of gravity on the $\omega - a_0(0)$ relation. Left: Curves for $\lambda=0.75$ and $\lambda=1$ with $\gamma=0$ (as in Fig. \ref{type_0_I}-Left) and $\gamma=1/160$. Right: Type-1 curves for $\lambda=1$ with four values of $\gamma$  starting from $\gamma=0$.
\label{grav_omega}
}
\end{center}
\end{figure}

\subsection{Numerical results}
Fixing the constant $\gamma > 0$,  the solutions can be constructed  by increasing progressively the amplitude of the
Proca field on the center, i.e.  $a_0(0)$.
 First it should be mentioned that no Type-0 solutions could be found in the gravitating case.
Attempts to deform a Type-0 solution by coupling it to gravity lead to the occurrence of a singularity at a finite value of the radial coordinate.
As a consequence,   no bifurcation of the type of Fig. \ref{type_0_I} could be found.
For all pairs of $\gamma, \lambda$ that we choose,
 the curves of gravitating Q-tubes are limited by  non-analytical configurations of the vector potential
when a critical value of the control parameter $a_0(0)$ is approached.
More precisely  the field $a_1 (x)$ reached a more and more pronounced  wall-shape at some $x=x_w$, the value $|a'_1(x_w)|$
becomes very large and likely tends to infinity.
The profile of such a behavior is shown in Fig. \ref{grav_profile}
for $\gamma = 1/40$;   it has $a_0(0) = 0.55$ and $x_w \approx 1.688$
 (note that the function  $a'_1(x)$ is still finite in this case:  $a'_1(x_w) \approx -5.0$
 but we truncated the figure for clarity). The metric functions are presented on the right  hand side of
 Fig.\ref{grav_profile}  through the combinations $xN'/N$, $xK'/K$ and $xL'/L-1$  characterizing
 the deviation of the metric potentials with respect to the Minkowski space-time,  and stressing the asymptotic power-law behavior. The Kasner powers can be read on the figure
 we find $a \approx 0.1773$, $c \approx -0.15$ and the relation $a+b+c-1$ is fulfilled within $10^{-4}$. 

\begin{figure}[t!]
\begin{center}
{\includegraphics[width=8cm]{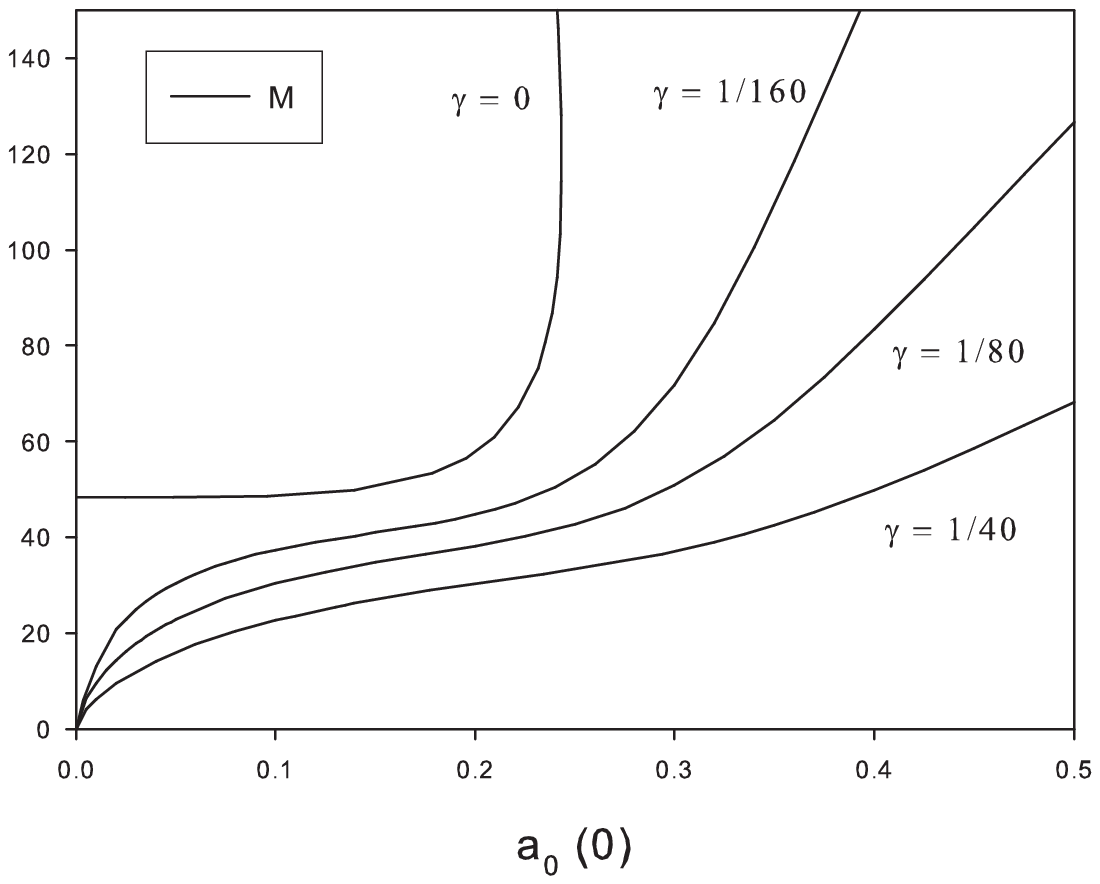}}
{\includegraphics[width=8cm]{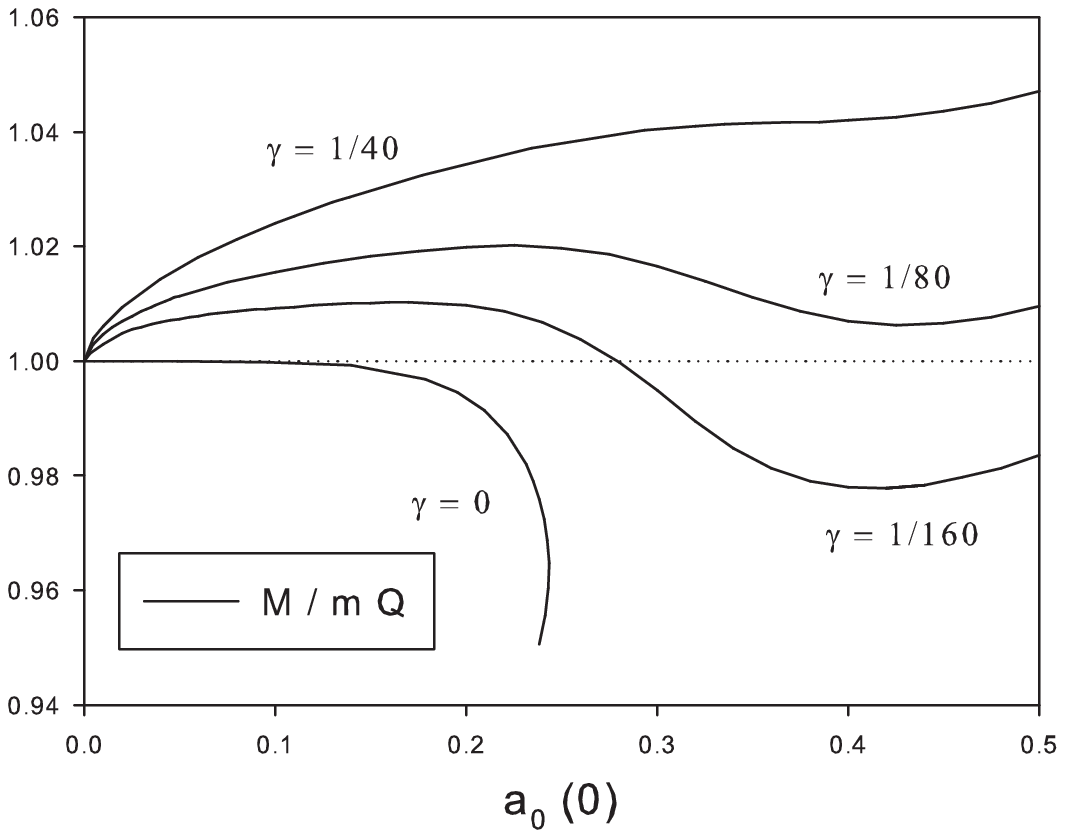}}
\caption{Left: The mass of the gravitating solutions as a function of $a_0(0)$ for $\lambda=1$ and four values of $\gamma$, starting from $\gamma=0$.
Right: The dependence of the ratio $M/Qm$ on $a_0(0)$ for the same parameters.
\label{grav_mass}
}
\end{center}
\end{figure}


Several features of gravitating Proca Q-tubes are illustrated in Figs. \ref{grav_omega} and \ref{grav_mass}. The influence of gravity on the Q-tube $\omega - a_0(0)$ relations is shown in Fig. \ref{grav_omega}. The singular configurations limiting  the gravitating solutions are symbolized by the bullets. Note however the similarity between the flat space type-0 curves and those of the localized (finite $Q$ and $M$) self-gravitating solutions with large $a_0 (0)$. It seems that when $\gamma >0$, the two branches merge into one continuous curve of localized solutions only.
The right hand side of Fig. \ref{grav_omega} reveals that when the gravitational parameter increases, gravitating
 solutions with $\omega > 1$ exist on a large domain of the control parameter $a_0(0)$.  This is not a violation of the $\omega<1$ bound, because of the time dilation in the Kasner spacetime represented by the $N(x)$ factor. Therefore, it is $\omega/N(x)$ which should be smaller than 1 throughout most space as is indeed the case. Note also Eq. (\ref{a0Asympt-Grav}) which does not impose a limit on the parameter $\omega$ in the self-gravitating case as Eq. (\ref{type_I}) does in the probe limit.

Another aspect of the gravitational effects on the Q-tubes is shown in Fig. \ref{grav_mass}. It presents the $a_0(0)$ dependence of the mass and stability parameter $M/(Qm)$.
 The left side of the figure shows in particular that the mass of the gravitating solutions vanishes in the limit $a_0(0) \to 0$
 (in a clear contrast with the probe limit) and that, for a fixed $a_0(0)$, the mass is generally much lower than without gravity. This is actually expected, due to the fact that gravity now makes the system ``more bound''. The right panel shows that stable self-gravitating Q-tubes are much less common than their flat space counterparts.

 As for the dependence on the parameter $\lambda$, generally it has a similar effect as in the probe limit, namely, the mass and charge increase with $\lambda$ but they have quite a small upper bound. Since the region of stability (i.e. when $M/Qm <1$) is much more limited when $\gamma>0$, we didn't perform here a full analysis of the $\lambda$-dependence of the self-gravitating Q-tube characteristics.

\subsection{Stability}
\begin{figure}[t!!]
\begin{center}
{\includegraphics[width=8cm]{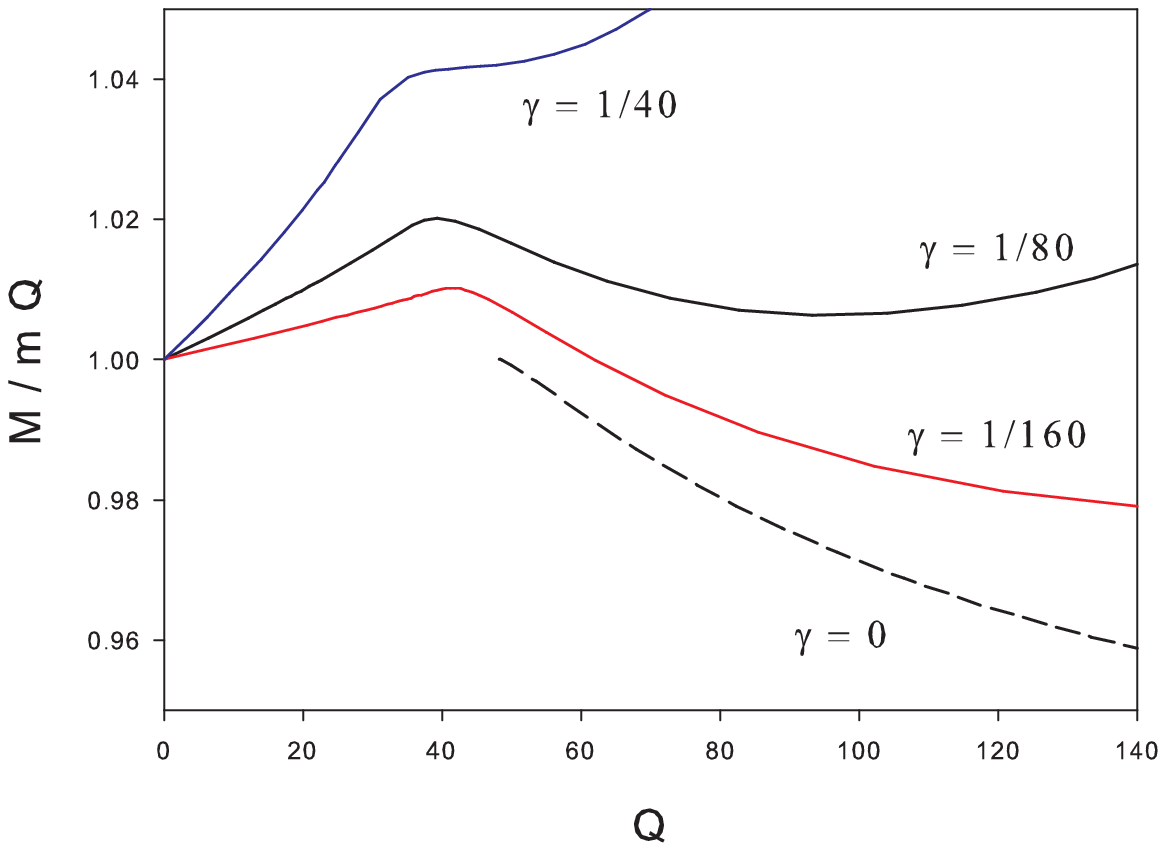}}
\includegraphics[width=8cm]{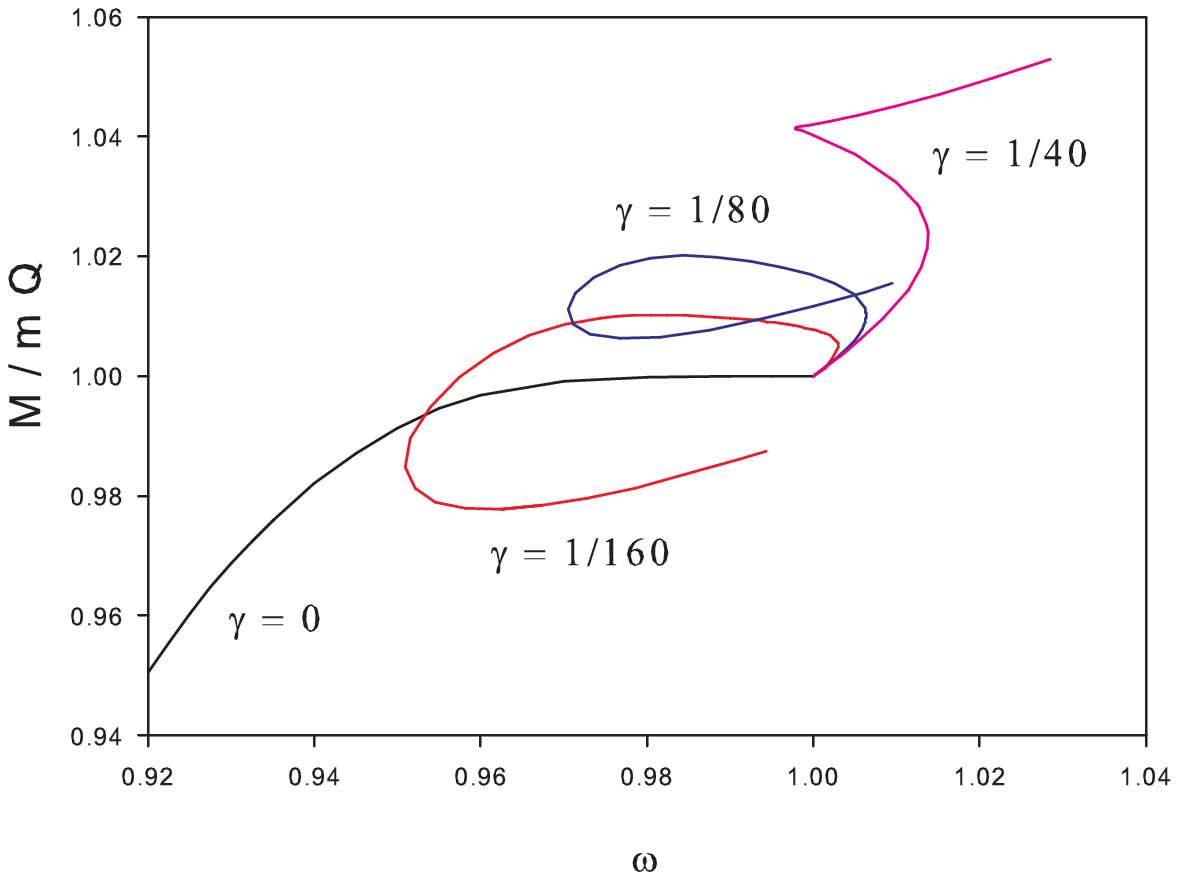}
\caption{The dependence of the stability ratio $M/Qm$ on the particle number $Q$ (left) and on $\omega$ (right) for four values of $\gamma$, starting from $\gamma=0$.
\label{Q_MoverQ}
}
\end{center}
\end{figure}

 In the presence of gravity a smaller number of particles is needed in order to form localized bound structures.
  Thus $Q$ is now much smaller too. This preference of small number of particles is reflected also in the fact that when $Q$ is increased the Q-tubes become unstable and ``prefer'' to undergo ``fission'' into a number of smaller stable tubes - see Fig. \ref{Q_MoverQ}. This phenomenon is common to scalar Q-balls \cite{Multamaki-Vilja2002,Verbin2007}, but here it is much more prominent.

\begin{figure}[b!!]
\begin{center}
{\label{interior_type_1}\includegraphics[width=8cm]{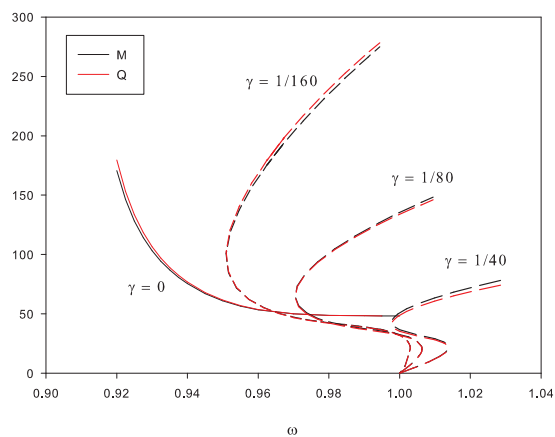}}
\end{center}
\caption{ The dimensionless mass and charge as a function of $\omega$ for four values of $\gamma$, starting from $\gamma=0$.
\label{M+Qvsomega}
}
\end{figure}

 From the right panel of Fig. \ref{grav_mass} and from Fig. \ref{Q_MoverQ} we see also that the stability ratio $M/mQ$ is often larger than 1 and the corresponding solutions are unstable. The larger (relatively) $\gamma$ values (above $\gamma \approx 0.01$)  yield solutions which are unstable throughout the whole branch. On the other hand the non-gravitating Q-tubes are always stable. In the middle there exists a small region where the solutions are stable for a certain interval of charge values or $a_0(0)$ values.

 Finally we present in Fig.\ref{M+Qvsomega} the $\omega$ dependence of the Q-tube mass and $U(1)$ charge for 4 values of the gravitational strength parameter $\gamma$. We notice a sharp difference in the behavior with respect to the analogous plot for Proca Q-balls - Fig. 6 of ref. \cite{Landea-Garcia2016}: As $\omega\rightarrow 1$ both $Q$ and $M$ vanish as in the spherically-symmetric case, but unlike that case, $Q$ and $M$ increase with increasing $\omega$ up to a maximal value of $\omega$ slightly larger than 1. Then $Q$ and $M$ increase significantly, while $\omega$ undergoes several oscillations, until the curves terminate at the maximal $a_0 (0)$. This is much unlike the spiral behavior of $Q$ and $M$ presented in \cite{Landea-Garcia2016}.


 \section{Conclusion}\label{Conclusion}
\setcounter{equation}{0}
In this paper we presented and analyzed the main properties of Proca Q-Tubes, that is cylindrically-symmetric localized solutions of a self-interacting massive vector field with a global conserved $U(1)$ charge. We found that the system is defined by 2 independent parameters, $\lambda/(m\nu^{1/2})$ and $\gamma=8\pi G m/\nu^{1/2} $. For a given pair of these parameters, there exists a family of ``ground state'' solutions defined by the central value of the vector field, $a_0 (0)$. These ``ground state'' solutions have a minimal number of nodes: $a_0 (x)$ has one node at some $x>0$, while $a_1 (x)$ has a node at $x=0$. We found several families of radially excited states with an increasing number of nodes and correspondingly increasing mass and charge.

These Q-tubes are stable in a large portion of parameter space if gravity is neglected. However, when gravity is taken into account it limits quite significantly the stability region. In physical terms, stable solutions seems to exist for small values of the gravitational strength $\gamma$ up to about 0.01, and a relatively small value of particle number per unit length as seen in Fig. \ref{Q_MoverQ}.
These results are the product of a first survey and a much more extensive study is needed in order to reveal more detailed aspects of these structures and to understand them in more depth. This study was limited to static solutions of the ``electric'' type whose geometry is asymptotically Kasner. A further study is required as to the question of the existence of ``magnetic'' solutions (static or stationary) and the possibility of asymptotically conic geometry. Preliminary results show that indeed some self-gravitating magnetic Proca tubes are asymptotically conic.


\end{document}